\RequirePackage{xcolor}
\documentclass[letterpaper, 10 pt, conference]{ieeeconf}  
\usepackage{float}
\usepackage{TLC}
\usepackage{xurl}
\usepackage{url}
\usepackage[normalem]{ulem}
\usepackage{setspace}
\let\labelindent\relax
\usepackage{enumitem}
\usepackage{mathtools}
\usepackage{gensymb}
\usepackage{amsmath}

\usepackage{epigraph}
\usepackage{shellesc}

\usepackage[style=ieee,
            url=false,
            date=year]{biblatex}

\theoremstyle{remark}

\title{A Winner-Takes-All Mechanism \\ for Event Generation}

\author{Yongkang Huo, Fuvio Forni, Rodolphe Sepulchre%
\thanks{The authors are with the University of Cambridge, Department of Engineering, Trumpington Street, CB2 1PZ, \texttt{\{yh415, \}@cam.ac.uk}.  R. Sepulchre is also with KU Leuven,
Department of Electrical Engineering (STADIUS),
KasteelPark Arenberg, 10,
B-3001 Leuven, Belgium,
\texttt{rodolphe.sepulchre@kuleuven.be}.}
\thanks{The research leading to these results has received funding from the European Research Council under the Advanced ERC Grant Agreement SpikyControl n.101054323. The work of Y. Huo was supported by the UK Engineering and Physical Sciences Research Council (EPSRC) grant 10671447 for the University of Cambridge Centre for Doctoral Training, the Department of Engineering. 
}%
}

\begin{document}

\maketitle
\thispagestyle{empty}
\pagestyle{empty}

\begin{abstract}
We present a novel framework for central pattern generator design that leverages the intrinsic rebound excitability of neurons in combination with winner-takes-all computation. Our approach unifies decision-making and rhythmic pattern generation within a simple yet powerful network architecture that employs all-to-all inhibitory connections enhanced by designable excitatory interactions. This design offers significant advantages regarding ease of implementation, adaptability, and robustness. We demonstrate its efficacy through a ring oscillator model, which exhibits adaptive phase and frequency modulation, making the framework particularly promising for applications in neuromorphic systems and robotics.
\end{abstract}

\section{Introduction}
Central pattern generators provide a bio-inspired framework for locomotion control in robotics and neuromorphic systems by autonomously generating robust, rhythmic motor patterns. In many robotic applications \cite{pazzaglia_balancing_2025,ijspeert_central_2008,linares-barranco_towards_2022,ijspeert_central_2008,zhang_online_2024,ijspeert_simulation_2005,pazzaglia_balancing_2025,abadia_neuromechanics_2025}, these approaches enable smooth gait generation, rapid adaptation to disturbances and varying terrains, and reduced computational overhead.

Traditional models, such as half-center oscillators \cite{brown_intrinsic_1911,brown_nature_1914,wilson_central_1961,marder_central_2001,schmetterling_neuromorphic_2024}, have long been favored for their simplicity in producing basic alternating patterns. However, although effective, these models lack a generalized framework that supports more complex and interactive central pattern generators.

To overcome this limitation, we propose a new design framework — \emph{rebound winner-takes-all}  — that extends and generalizes the half-center oscillator 
paradigm. Our approach is based on two fundamental mechanisms.
The first is the 
\emph{winner-takes-all} computation \cite{majani_k-winners-take-all_1988,oster_computation_2009,maass_computational_2000}, a classical paradigm for neural computation with inhibitory networks. The basic concept is that inhibitory networks can select and memorize the ``dominant'' input out of a complex family of competing input signals. The second mechanism is 
\emph{rebound excitability} \cite{rush_potassium_1995},  a neuronal behavior characterized by a post-inhibitory rebound, that is, 
the ability to trigger an event response when inhibition is ``released''. The rebound behavior is a key feature of central pattern generators \cite{brown_intrinsic_1911,brown_nature_1914,wilson_central_1961,marder_central_2001}, which can robustly produce autonomous sequences of events in the presence of uncertain cellular and synaptic dynamics \cite{morozova_reciprocally_2022}. 

By leveraging the intrinsic rebound dynamics of neurons in combination with winner-takes-all computation, our mechanism selectively enhances neuron activation in response to both sensory inputs and internal excitations. This not only ease the design and enables the generation of richer and more adaptable rhythmic outputs, but also preserves the simplicity inherent in the half-center oscillator.

A key design advantage is its modularity and ease of control. The network is built upon all-to-all inhibitory connections, augmented with tunable excitatory interactions. This architecture allows for straightforward programming of diverse rhythmic patterns, facilitates rapid adaptation to changing inputs, and ensures robustness even when the physical realization of neuronal rebound properties varies.

In this paper, we focus on a ring oscillator as a concrete example to demonstrate the properties and design advantages of the rebound winner-takes-all network. The underlying principles are readily extendable to more complex network architectures, paving the way for advanced neuromorphic systems and robotic controllers with improved adaptability and performance.

{The paper is organized as follows. 
Neural modeling and the key property of rebound excitability are discussed in Section \ref{sec:rebound}. Rebound excitability is 
then used to construct a half-center oscillator in Sections \ref{sec:HCO}. The two 
fundamental mechanisms rebound excitability and winner-takes-all are combined in Section \ref{sec:rwta}, to construct ring oscillators, that is, robust reactive periodic pattern generators. Conclusions follow. The code for reproducing all plots will be available at \url{https://github.com/Huoyongtony/Rebound-winner-takes-all}}

\section{Rebound Excitability}\label{sec:rebound}
\begin{figure}[htbp]
    \centering
    \includegraphics[width=1\columnwidth]{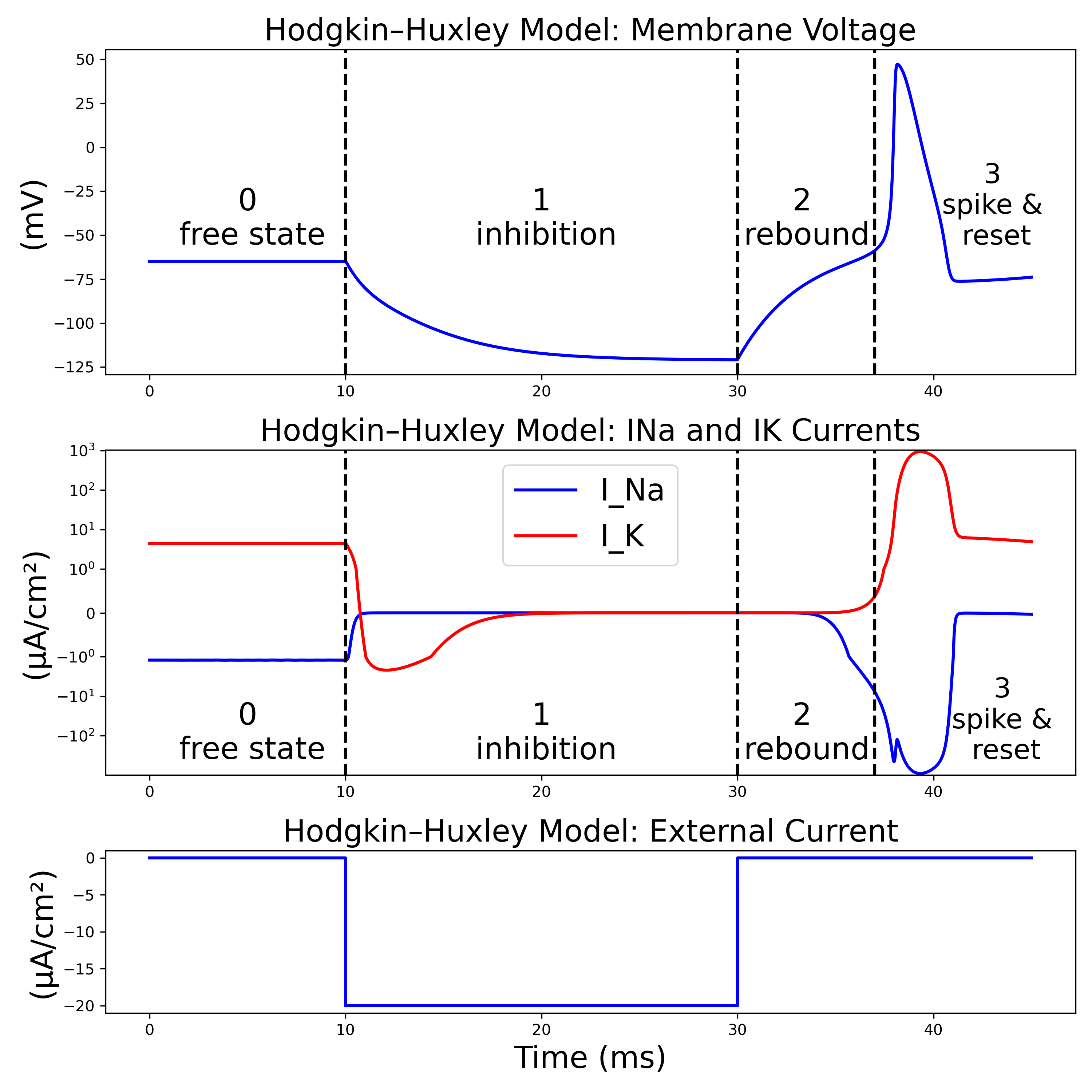}
    \caption{Rebound spiking in HH model.}
    \label{fig:rebound_spike}
\end{figure}

Rebound excitability is the property of a neuron to generate a spike or burst following a period of inhibition. For example, a neuron that remains silent under a constant input current \(u_0\) will emit a spike or burst when the input is briefly reduced and then restored to \(u_0\). Fig.\,\ref{fig:rebound_spike} illustrates this behavior through simulations of the Hodgkin–Huxley neuron model.

{ 
Rebound excitability of the Hodgkin–Huxley neuron model arises from the interplay of fast inward (depolarizing) currents $I_\text{Na}$ and slow outward (polarizing) currents $I_\text{K}$ that provide delayed feedback. We can express the membrane dynamics of a Hodgkin-Huxley rebound neuron in a simplified form as follows:


}

\begin{definition}[Hodgkin-Huxley rebound neuron]\label{eq:membrane}
\begin{align*}
    C\frac{dV}{dt} &= -I_\text{Leak} - I_\text{Na} - I_\text{K}\\ 
    &\quad + I_{\rm ext}(t)+\sum_i I_{\rm syn\_i}(V_i)
\end{align*}
{ 
Here, \(V\) is the membrane potential. $I_\text{Leak}=-g_L(V-E_{L})$ is the leakage current. $I_\text{Na}=-G_\text{Na}(V-E_{Na})$ and $I_\text{K}=G_\text{K}(V-E_K)$ with dynamic conductances $G_\text{Na}$ and $G_\text{K}$. \(I_{\rm ext}(t)\) is the external input current. The full detail of the dynamics is given in the appendix.

\(\sum_i I_{\rm syn\_i}(V_i)\) denotes the synaptic input from other neurons. In this paper we adopt the simplified model for the synaptic current \(I_{\rm syn\_i}(V_i)\) of \cite{ribar_neuromorphic_2021-2}, which has the following dynamics:
\begin{equation*}\label{eq:syn}
\begin{aligned}
\quad I_{\rm syn\_i}&=\sigma(V_f-V_\text{th}), \quad \tau \dot{V_{f}} = V_\text{i} - V_f\\[1mm]
\sigma(V) &= \frac{g_\text{syn\_i}}{1 + \exp\bigl(- \alpha \cdot V \bigr) }
\end{aligned}
\end{equation*}
where \(\tau\), \(\alpha\), \(g_\text{syn\_i}\) and \(V_\text{th}\) are constant parameters and all the \(V_\text{i}\) represent the membrane voltages of othe neurons. \(g_\text{syn\_i}\) is the synaptic strength of a synapse. \(g_\text{syn\_i}\) is positive for an excitation synapse and is negative for an inhibition synapse. \qedhere}
\end{definition}

$I_\text{Na}$ acts as a positive feedback component that pulls the membrane voltage up to $E_{Na}$, whereas $I_\text{K}$ serves as a negative feedback component that brings the membrane voltage back to the resting potential.

The rebound mechanism unfolds in four stages, as shown in Fig.\,\ref{fig:rebound_spike}:
\begin{enumerate}[start=0]
    \item \textbf{Free stage:} During the free stage, the neuron receives a constant external current and remains silent.
    \item \textbf{Inhibition:} During an inhibitory pulse, the membrane potential is driven far below the resting potential. In this state, both the fast-inward and slow-outward currents are inactivated.
    \item \textbf{Rebound:} Upon removal of inhibition (i.e., when \(I_{\rm ext}(t)\) returns to baseline), the $I_\text{Na}$ current is rapidly activated, while the $I_\text{K}$ current is activated with a slight delay, leading to an overshoot in the membrane voltage.
    \item \textbf{Spike Initiation and Reset:} If the overshoot is large enough, the threshold to trigger a spike is reached. The spike is brief because it is quickly counteracted by $I_\text{K}$, which rapidly brings the membrane potential back to the resting level.
\end{enumerate}


\section{Half-Center Oscillator}\label{sec:HCO}
Central pattern generators are neural networks capable of generating rhythmic motor outputs in the absence of sensory input or descending commands, while still adapting to external inputs. In this section, we detail the simplest central pattern generator model, the half-center oscillator architecture.

\begin{figure}[thbp]
    \centering
    \includegraphics[width=1\columnwidth]{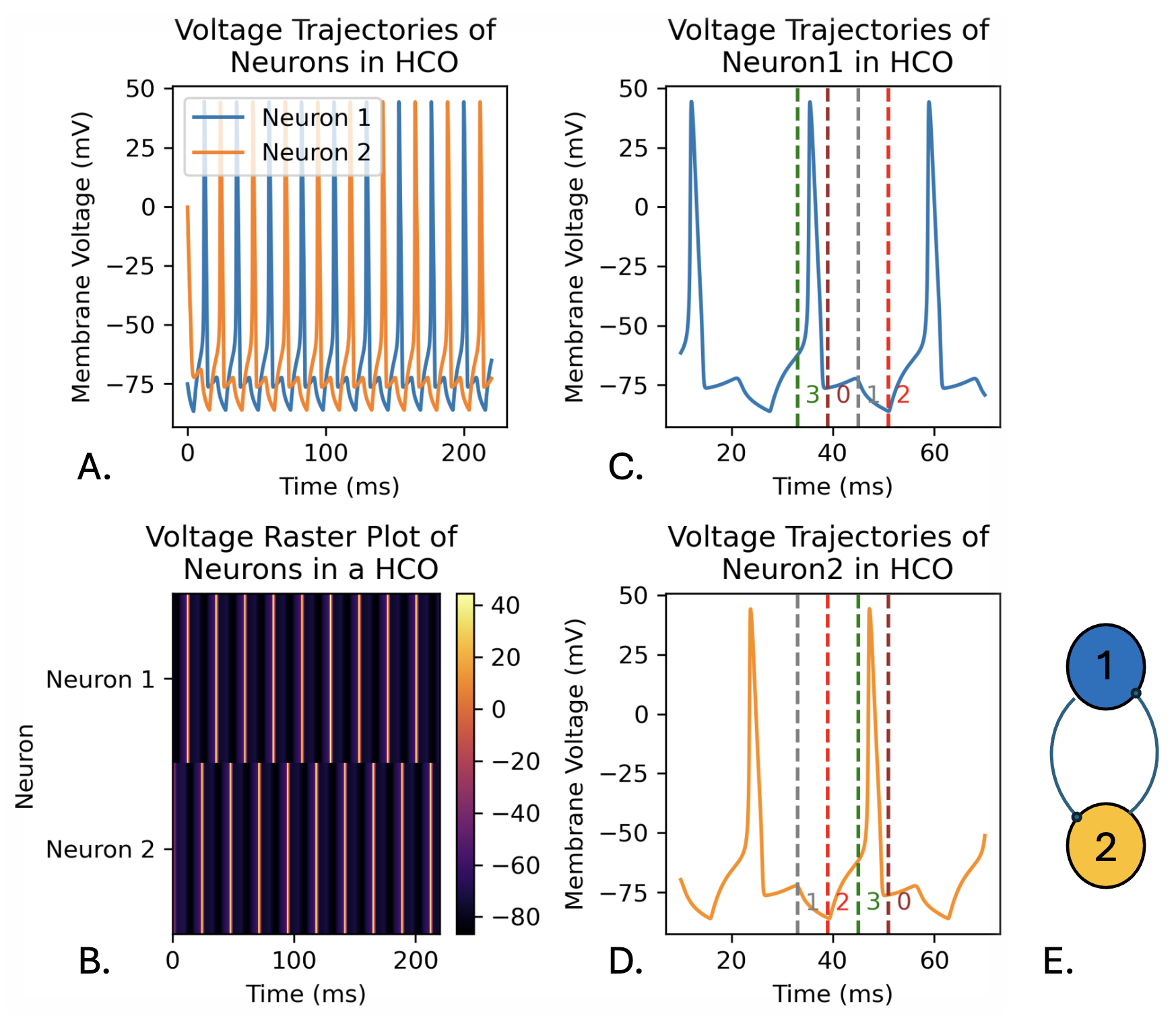}
    \caption{HCO. The parameters for inhibitory synapses are [$g_\text{syn}=10,\tau=1,V_\text{th}=-65,\alpha=1.5$].}

    \label{fig:HCO}
\end{figure}

The half-center oscillator usually consists of two neurons (or neural populations) with reciprocal inhibitory connections, as shown in Fig.\,\ref{fig:HCO}(E). Synaptic connections are inhibitory, so \(g_\text{syn}\) is negative. Typically, inhibitory connections in a half-center oscillator are fast (e.g., with \(\tau=1\)). Fig.\,\ref{fig:HCO}(C) and (D) illustrate details of this mechanism. When neuron 1 is in the spiking stage (3), it inhibits neuron 2 via the inhibitory synapse, placing neuron 2 in the inhibition stage. Shortly after spiking, neuron 1 returns to the free stage, releasing neuron 2 from inhibition so that it can rebound. After rebounding, neuron 2 enters the spiking stage and inhibits neuron 1 through the inhibitory synapse. Finally, neuron 2 returns to the free stage after spiking, allowing neuron 1 to begin its rebound. This cycle repeats, generating the oscillatory pattern shown in Fig.\,\ref{fig:HCO}(A) and (B). This mechanism is well described in the literature, notably in \cite{marder_central_2001} and \cite{dirk_central_2015}.

\section{Rebound Winner-Takes-All network}\label{sec:rwta}

We define a rebound winner-takes-all network as an all-to-all inhibitory network of rebound neurons. There are four rules in the rebound winner-takes-all network:
\begin{itemize}
    \item Only one neuron can be active at any time due to the all-to-all inhibition. The active neuron is the `winner'.
    \item At the rebound stage, the system's dynamics exhibit winner-takes-all behavior: the neuron receiving the highest stimulus or having the highest initial voltage will emerge as the winner (active). 
    \item The neuron that won in the previous cycle will not win again in the next cycle, unless an extremely large external input is applied. This is because the winning neuron enters the free stage after spiking (Figure \ref{fig:rebound_spike}), while all other neurons remain in the rebound stage.
    \item After one neuron is activated, all other neurons transition to the rebound stage due to all-to-all inhibition, ensuring that eventually one of them becomes active.
\end{itemize}

The simplest example of a rebound winner-takes-all network is the half-center oscillator, which is an all-to-all inhibitory network composed of two rebound neurons. One can verify that the half-center oscillator adheres to the three aforementioned rules: (1) only one neuron is active at a time; (2) the neurons spike alternately, meaning that the previous winner does not spike in the subsequent cycle; (3) the oscillatory pattern is recurrent, both in the presence and absence of external inputs.

Our proposed rebound winner-takes-all network has the same topology of more traditional winner-takes-all networks \cite{majani_k-winners-take-all_1988,oster_computation_2009,maass_computational_2000}, where multiple units compete to become the single “winner” or dominant active unit.
The only difference is the type of neuron used in the network: we use rebound neurons (for instance, Hodgkin–Huxley neurons) instead of Hopfield neurons.

In terms of behavior, both networks allow at most one neuron to be active at any time, with the neuron receiving the largest stimulus emerging as the winner. However, in our framework, the neuron that won in the previous cycle does not win again in the next cycle. In contrast, in the classical winner-takes-all network, every neuron has a chance to win the competition in each cycle. For winner-takes-all networks with auto-reset \cite{kohonen_physiological_1993,grossberg_nonlinear_1988}, a recurrent sequence of events can be produced; however, such recurrence must be driven by external inputs (the sequence stops in the absence of external input), whereas the recurrent behavior in the rebound winner-takes-all network is an intrinsic property driven by the rebound current.

\begin{figure}
    \centering
    \includegraphics[width=1\columnwidth]{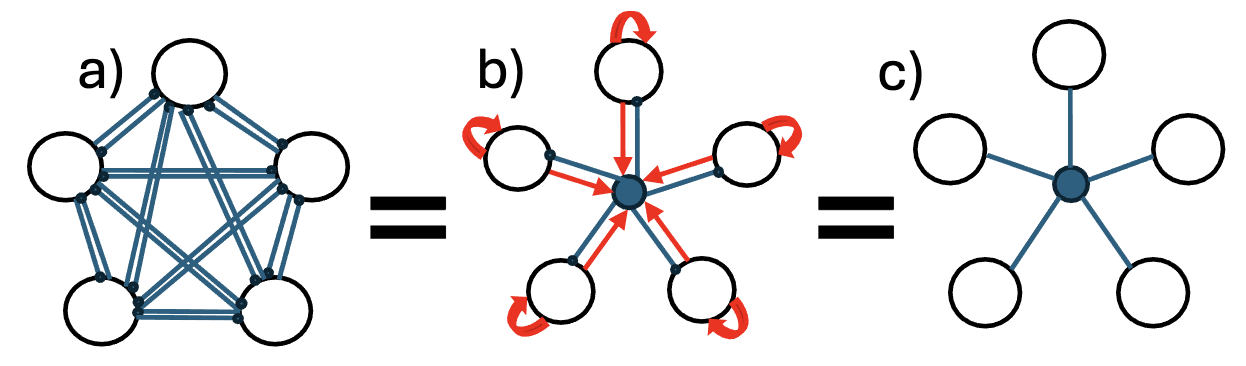}
    \caption{Representations of an all-to-all inhibitory network}
    \label{fig:inhibition}
\end{figure}

Fig.\,\ref{fig:inhibition}(a) illustrates a fully connected winner-takes-all topology in a network of five neurons. Because a fully connected network of \(N\) neurons requires \(N^2\) connections, it is neither efficient nor biophysically plausible. A common simplification in winner‐take‐all network models is to replace fully connected (all-to-all) inhibitory circuitry with a single 'center' inhibitory neuron (or group) that collects excitatory signals from all competing units and feeds back inhibition uniformly, as shown in Fig.\,\ref{fig:inhibition}(b). Although mathematically equivalent, this simplification is not only computationally convenient, but also biologically plausible. Several studies \cite{kaski_winner-take-all_1994,coultrip_cortical_1992,lazzaro_winner-take-all_1988,douglas_neuronal_2004,douglas_canonical_1989} have provided evidence for similar architectures in the cortex and other areas of the brain, where inhibitory interneurons, such as parvalbumin-expressing basket cells, can combine input from large populations of excitatory neurons and provide broad, nonspecific inhibition. In the present paper, we focus exclusively on the realization depicted in (a); for notational simplicity, we represent the all-to-all inhibitory network as shown in Fig.\,\ref{fig:inhibition}(c).

\begin{figure*}
    \centering
    \includegraphics[width=\textwidth,height=6cm]{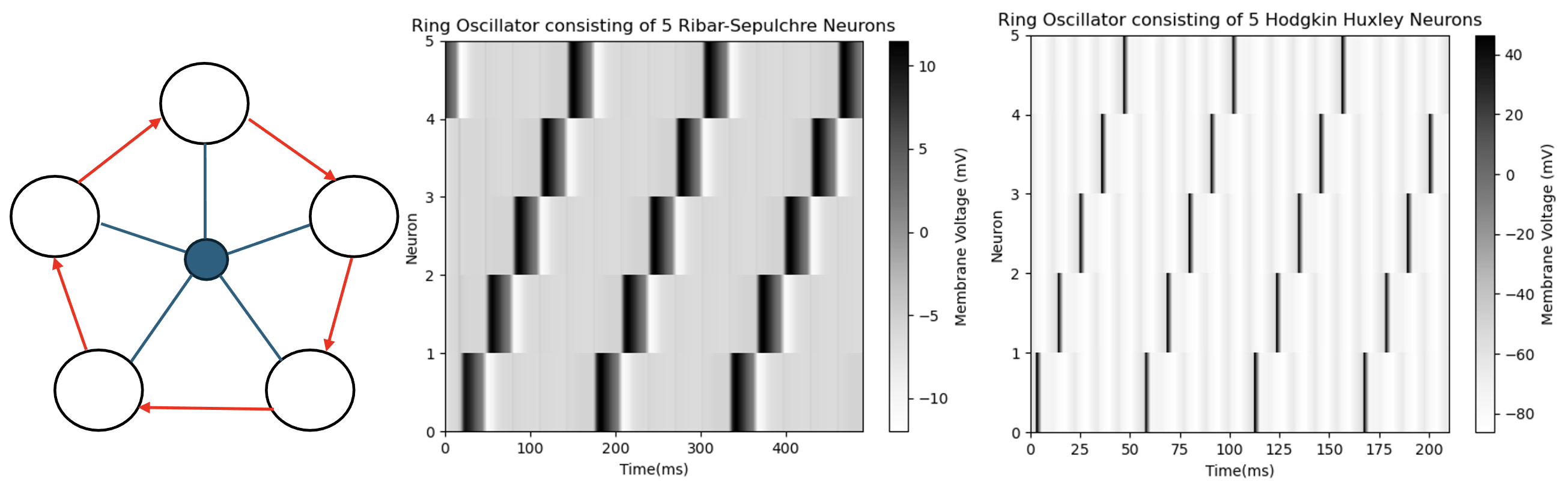}
    \caption{On the left is the notation for the rebound winner-takes-all ring oscillator. On the right are two raster plots of membrane voltages for two ring oscillators, each composed of different types of neurons. For the ring of Hodgkin-Huxley neurons the parameters for inhibitory synapses are [$g_\text{syn}=-10,\tau=1,V_\text{th}=-65,\alpha=1.5$] and the parameters for excitatory synapses [$g_\text{syn}=0.5,\tau=5,V_\text{th}=-65,\alpha=1.5$]. For the ring of Ribar-Sepulchre neurons the parameters for inhibitory synapses are [$g_\text{syn}=-5,\tau=0.1,V_\text{th}=-4,\alpha=2$] and the parameters for excitatory synapses [$g_\text{syn}=0.3,\tau=60,V_\text{th}=-4,\alpha=2$].}
    \label{fig:ring}
\end{figure*}

\begin{figure*}
    \centering
    \includegraphics[width=\textwidth,height=9cm]{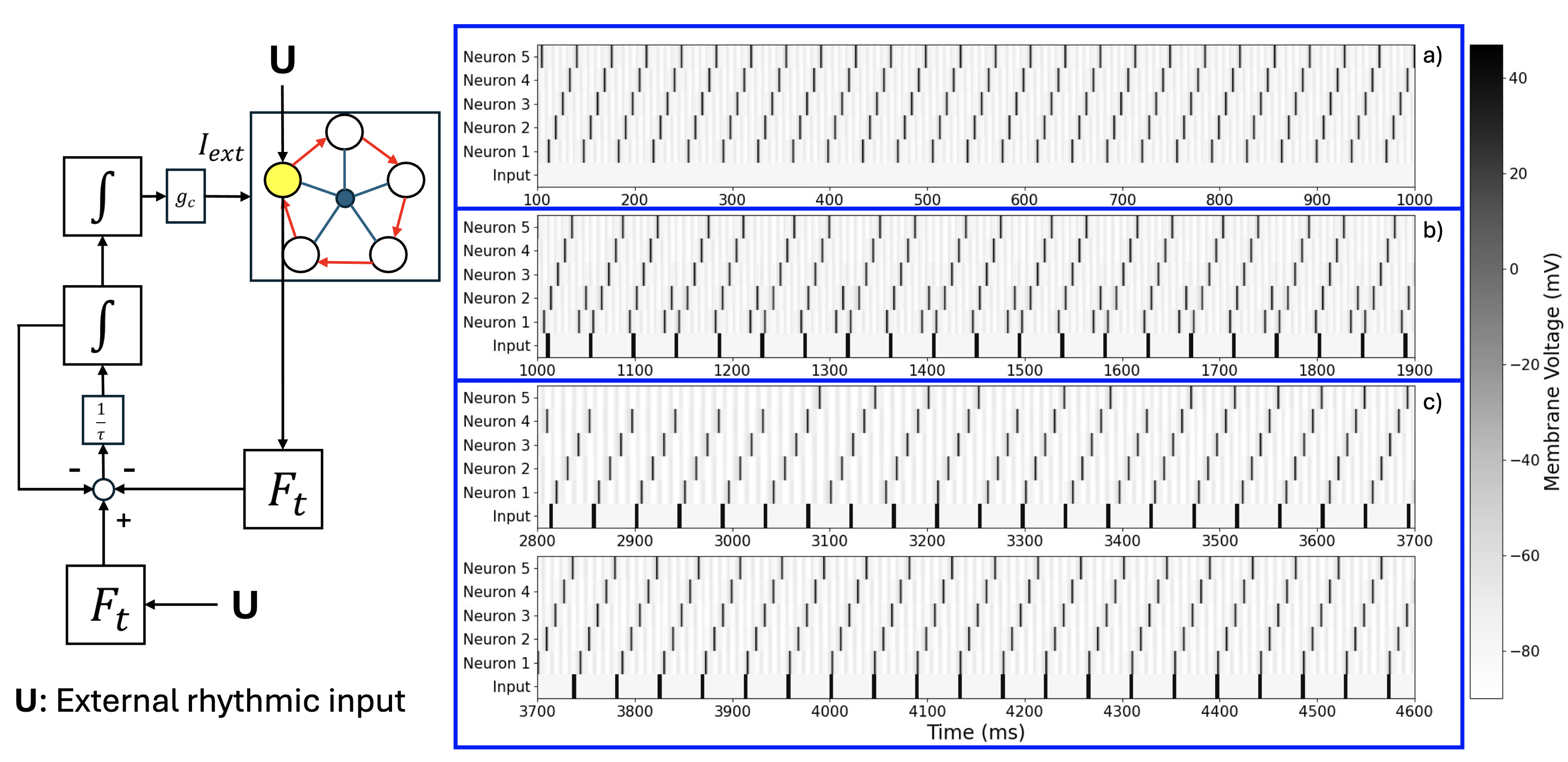}
    \caption{Synchronizing a ring oscillator composed of 5 Hodgkin–Huxley neurons to an external rhythmic input. Parameters for inhibitory synapses [$g_\text{syn}=-15,\tau=0.1,V_\text{th}=-65,\alpha=1.5$]. Parameters for excitatory synapses [$g_\text{syn}=10,\tau=0.1,V_\text{th}=10,\alpha=1.5$]. a) Endogenous rhythm of the ring. b) The ring being entrained by the external input. c) Adaptive controller is applied together with the external rhythmic input. }
    \label{fig:adap_entrain}
\end{figure*}

\section{Rebound Winner-Takes-All \\Central Pattern Generator}

\subsection{Endogenous Behavior}

In this section, we construct a robust ring oscillator from a rebound winner-takes-all network by introducing excitatory synapses.
The excitatory synapses follow the dynamics in Definition~\ref{eq:membrane}, but with a positive \(g_\text{syn}\). Excitatory synapses are shown in red in Fig.\,\ref{fig:ring}.

We start with the analysis of the endogenous behavior of a ring oscillator. Assume that one neuron has just emitted an event; the excitatory synapse slightly excites the next neuron, establishing a preference for the subsequent winner. In the absence of external inputs, the neurons in the ring are activated sequentially, forming a ring oscillator. 

In this example, we use five Hodgkin–Huxley neurons arranged in a ring. The parameters for excitatory connections are \(\tau=5\), \(\alpha=1.5\), \(\text{th}=-65\), and gain \(=0.5\); for the inhibitory connections, they are \(\tau=1\), \(\alpha=1.5\), \(\text{th}=-65\), and gain \(=-10\). The voltage trajectory raster plot for this ring oscillator is shown in Fig.\,\ref{fig:ring}.

Additionally, this framework works with other types of rebound neurons. In Fig.\,\ref{fig:ring}, we also present a voltage trajectory raster plot for a ring oscillator comprising five Ribar-Sepulchre rebound neurons \cite{ribar_neuromorphic_2021}.

\subsection{Exogenous Behavior - synchronization via pulses}\label{sec:exo_sync}

In the presence of an external input, the phase of the ring oscillator can be controlled through entrainment. The external input represents the influence of a synaptic current external to the ring network.

{
We consider an isolated external input acting on one neuron of the network. We can assume that this input is weak relative to the nodal ($I_\text{Na}$ and $I_\text{K}$) and inhibitory synaptic currents. Such an external input will not destroy the endogenous rhythm. Instead, it will only control the phase of the pattern, that is, the time intervals between successive events. This is known as entrainment in biology. This effect will be used for phase control in \ref{sec:control} and is shown in Fig.\ref{fig:adap_entrain}(b).

The effect of a weak external pulse is rather intuitive: we can understand its effect by understanding external input applied to a single neuron at the four different stages.

\begin{enumerate}[start=0]
    \item \textbf{Free stage:} In this network, a neuron enters the free stage if it was the winner in the previous round, meaning that all other neurons are in the rebound stage. Since we assume that the external input is weaker than \(I_{\rm Na}\) in neurons in the rebound stage, a neuron in the free stage cannot win the current round. Therefore, external input applied during the free stage has no effect on the ring oscillator.
    \item \textbf{Inhibition:} External input at this stage alters the membrane voltage, thereby setting the initial voltage for the rebound stage. According to Rule 2, the neuron that begins with a higher initial voltage will win if no additional external input is provided during the rebound stage. Because the external input is assumed to be much weaker than the inhibitory synaptic current, the neuron remains in the inhibition stage until the inhibitory synapse is turned off. Thus, input during the inhibition stage can be used to bias the selection of the next winner.
    \item \textbf{Rebound:} During the rebound stage, external inputs function similarly to those in traditional winner-takes-all networks: they modify the neuron's membrane voltage and thereby influence the selection of the winner. Additionally, external input affects the membrane's charge-up speed, which can advance or delay the occurrence of an event.
    \item \textbf{Spike Initiation and Reset:} This stage is dominated by $I_\text{Na} $ and $I_\text{K}$. Since the external input is much weaker than these currents, it has negligible effect on the network during this stage.
\end{enumerate}

}

\subsection{Exogenous Behavior - frequency adaptation via bias}\label{sec:exo_freq}

The frequency of the ring oscillator can also be controlled via external inputs. 
Frequency control refers to a global modulation of the network by adjusting the average external input $I_\text{ext}$ to the neurons. Modulating this average input does not confer any competitive advantage to any particular neuron; hence, it does not alter the order of events. Instead, it affects the overall level of inhibition, specifically, the initial membrane potential in the rebound stage, which, in turn, influences the timing of the rebound event and the rhythm frequency of the network. In the later section\ref{sec:control}, we demonstrate how this property can be used for phase and frequency control and this will be shown in Fig.\ref{fig:adap_entrain}(c).



\subsection{An example of pattern control}\label{sec:control}

We aim to synchronize the first neuron of the ring oscillator with an external rhythmic input. To simulate variability, Gaussian noise with a mean of 0 and a variance of 0.1 is added to each neuron as external input.
{
The ring's frequency is monotonically modulated by the average external current generated by an adaptive frequency controller. Looking at Figure \ref{fig:adap_entrain}, the adaptive controller takes as inputs the voltage of a neuron, \(u_v\), and the external rhythmic signal, \(u_r\), and computes the average external current \(I_\text{apply}\) as follows
\begin{align}
    u &= F_t(u_r)-F_t(u_v),  \nonumber \\
    \dot{e} &= \frac{u-e}{\tau}, \nonumber \\
    \dot{I}_\text{ext} & = g_c\cdot e \nonumber
\end{align}
where $F_t$ is an event detector such that $$F_t(u)=\sum_n\delta(t-t_n), \{t_n \; |\; u(t_n)=-40,\; u(t_n^-)<-40\}.$$ It detects events in the inputs based on when it cross the threshold $-40$ from below and converts them into train of unit pulses. The reason for this is to unify the events such that $e$ represents the differences in event frequencies. The event detector used in the current paper is just for simplicity while there are physical realizations of an event detector such as a spiking neuron, which might be explored in the future studies. The pulses is then filtered by a first order low-pass filter with time constant $\tau=250$, and integrated with a gain 
$g_c=\frac{2}{250}$.

For the ring synchronization,
we pass the external rhythmic input \(u\) through \(\sigma(u)\) with parameters $g_\text{syn}=2,V_\text{th}=-45,\alpha=1.5$ to generate the synaptic input to the neuron we wish to synchronize. As discussed in Section~\ref{sec:exo_sync}, when the input is applied at the right stages, this input forces neuron 1 to be either the next or the current winner, effectively inducing a phase jump. These phase jumps maintain the phase of the ring oscillator close to that of the external rhythmic input, minimizing large phase differences. As a result, the adaptive frequency controller primarily needs to match the frequency and minor phase discrepancies, thereby converging faster.
}
Fig.\,\ref{fig:adap_entrain}(a) shows the initial endogenous rhythm of the ring. Fig.\,\ref{fig:adap_entrain}(b) shows the ring being entrained by the external input, with frequent phase jumps evident due to a mismatch in frequency. In Fig.\,\ref{fig:adap_entrain}(c), the adaptive controller is applied together with the external rhythmic input. Initially, phase jumps are still present due to the frequency mismatch, but as the adaptive controller adjusts the average external current, the frequency of the ring becomes matched with that of the external input and phase jumps cease.

\section{Conclusion}
In this paper, we introduced a novel framework for central pattern generator design based on the rebound winner-takes-all mechanism, which seamlessly integrates decision-making with rhythmic pattern generation. By leveraging the intrinsic rebound excitability of neurons and incorporating winner-takes-all computation, our framework achieves robust and adaptable rhythmic output through a straightforward design that employs all-to-all inhibitory connections combined with flexible excitatory interactions.

The key design advantages of our approach include its ease of implementation, scalability, and inherent robustness, which together facilitate rapid adaptation to external stimuli and varying operational conditions. Our demonstration using a ring oscillator model highlights the framework's ability to extend traditional half-center oscillator architectures by providing enhanced control over phase and frequency dynamics. Future work will explore a more quantitative analysis of the behaviors of these networks (e.g., proving the four properties), scaling the network to more complex topologies, and implementing the design in hardware, thereby further validating its potential for real-world neuromorphic and robotic applications.

\section{Appendix}

\subsection{Hodgkin–Huxley model}
In definition \ref{eq:membrane} we have defined the membrane dynamics of the Hodgkin-Huxley model. Here we will give the full details of the dynamics of conductances.
\begin{align}
G_\text{Na}&=g_{Na}\,m^3h\, \nonumber\\[1mm]
G_\text{K}&=g_K\,n^4\,(V-E_K)-g_L,\,\nonumber\\[1mm]
\dot{x} &= \frac{x_\infty(V)-x}{\tau_x(V)},\quad x\in\{m,h,n\}, \nonumber
\end{align}
\[
x_\infty(V)=\frac{\alpha_x(V)}{\alpha_x(V)+\beta_x(V)},
\]
$$
\tau_x(V)=\frac{1}{\alpha_x(V)+\beta_x(V)}.
$$

\[
C=1\,\mu\mathrm{F/cm}^2,\; g_{Na}=120\,\mathrm{mS/cm}^2,
\]
\[
g_K=36\,\mathrm{mS/cm}^2,\; g_L=0.3\,\mathrm{mS/cm}^2,
\]
\[
E_{Na}=50\,\mathrm{mV},\; E_K=-77\,\mathrm{mV},\; E_L=-54.387\,\mathrm{mV}.
\]
$$
\small
\begin{aligned}
\alpha_m(V) &= \frac{0.1\,(V+40)}{1-\exp\left(-\frac{V+40}{10}\right)}, \quad
\beta_m(V) = 4\,\exp\left(-\frac{V+65}{18}\right), \\
\alpha_h(V) &= 0.07\,\exp\left(-\frac{V+65}{20}\right), \quad
\beta_h(V) = \frac{1}{1+\exp\left(-\frac{V+35}{10}\right)}, \\
\alpha_n(V) &= \frac{0.01\,(V+55)}{1-\exp\left(-\frac{V+55}{10}\right)}, \quad
\beta_n(V) = 0.125\,\exp\left(-\frac{V+65}{80}\right).
\end{aligned}
$$

\subsection{Ribar-Sepulchre model \cite{ribar_neuromorphic_2021}}
\begin{equation*}
\begin{aligned}
\dot{V}_1 &= \frac{V - V_1}{30}, \quad \dot{V}_2 = \frac{V - V_2}{60},\\[1mm]
\dot{V} &= -0.5\,V + 2\,\tanh\bigl(V-1\bigr) - 2\,\tanh\bigl(V_1-1\bigr) \\
&\quad - \tanh\bigl(V_1+2\bigr) + u,
\end{aligned}
\end{equation*}
where:
\begin{itemize}
    \item \(V\) is the membrane voltage,
    \item \(u\) is the external current input (including synaptic inputs).
\end{itemize}

\printbibliography

\end{document}